\documentclass{PoS}

\title{Composite flavor-singlet scalar in twelve-flavor QCD}

\ShortTitle{Composite flavor-singlet scalar in twelve-flavor QCD}

\author{
Yasumichi Aoki$^a$,
Tatsumi Aoyama$^a$,
Masafumi Kurachi$^a$,
Toshihide Maskawa$^a$,
Kohtaroh Miura$^a$,
Kei-ichi Nagai$^a$,
Hiroshi Ohki$^a$, 
Enrico Rinaldi$^b$\thanks{Present address: Lawrence Livermore National Laboratory, Livermore, California 94550, USA},
Akihiro Shibata$^c$,
Koichi Yamawaki$^a$,
and 
\speaker{Takeshi Yamazaki}$^a$

\hspace*{55mm} (LatKMI Collaboration) 
\\

$^a$
Kobayashi-Maskawa Institute for the Origin
of Particles and the Universe (KMI), Nagoya University, Nagoya
464-8602, Japan \\
$^b$ 
Higgs Centre for Theoretical Physics, SUPA, School of Physics and Astronomy, 
University of Edinburgh, Edinburgh EH9 3JZ, UK \\
$^c$
Computing Research Center, High Energy Accelerator Research Organization (KEK), 
Tsukuba 305-0801, Japan
\\
E-mail: \email{yamazaki@kmi.nagoya-u.ac.jp}
}

\abstract{
We report the calculation of the flavor-singlet scalar in
the SU(3) gauge theory with the degenerate twelve fermions in
the fundamental representation using a HISQ-type action 
at a fixed $\beta$.
In order to reduce the large statistical error coming from
the vacuum-subtracted disconnected correlator, we employ
a noise reduction method and a large number of configurations.
We observe that the flavor-singlet scalar is lighter than the pion 
in this theory from the calculations with
the fermion bilinear and gluonic operators.
This peculiar feature is considered to be due to 
the infrared conformality of this theory,
and it is a promissing signal for a walking technicolor,
where a light composite Higgs boson is expected to emerge by
approximate conformal dynamics.
}

\FullConference{31st International Symposium on Lattice Field Theory - LATTICE 2013\\
		July 29 - August 3, 2013\\
		Mainz, Germany}

\begin{document}

\section{Introduction}

Higgs boson was discovered at LHC experiment in the last year,
and its mass has also been already measured, which is about 125 GeV.
While the current experimental data insists that Higgs boson is 
consistent with the elementary particle in the Standard model, 
there still remains a possibility of a composite particle 
in a strongly coupled gauge theory.
One of such attractive models is 
the walking technicolor~\cite{Yamawaki:1985zg},
where a light composite flavor-singlet scalar is expected as 
a pseudo Nambu-Goldstone (NG) boson of the approximate conformal symmetry.
Phenomenologically, a consistency of a composite Higgs boson 
with the current LHC data was discussed in Refs.~\cite{Matsuzaki:2012mk,Matsuzaki:2012xx,Matsuzaki:2013fqa}.
Thus, the most urgent theoretical task 
is to check the existence of such a light flavor-singlet scalar state
from first principle, lattice gauge calculation. 
(For reviews on the lattice studies in search 
for candidates for the walking technicolor, 
see~\cite{Neil:2012cb,Giedt:2012hg,Kuti:2013xyz} and references therein.)

In this study, we calculate
the flavor-singlet scalar state ($\sigma$)
in the SU(3) gauge theory with the fundamental twelve fermions
($N_f = 12$ QCD),
which was studied by several groups.
In our previous study~\cite{Aoki:2012eq} of this theory,
it is found that $m_f$ dependence of meson masses
is consistent with the one expected in the conformal theory.
We aim to investigate the properties of $\sigma$ in this theory, 
especially whether $\sigma$ is lighter or not, because
$\sigma$ in the walking technicolor is expected to have 
similar properties due to approximate conformal symmetry.
Thus, this work is regarded as a pilot study for more attractive
theories, such as the $N_f = 8$ QCD that was found to be a 
candidate of the walking technicolor in our work~\cite{Aoki:2013xza}.

The calculation of $\sigma$ state is one of the most 
challenging calculations in the lattice simulation
due to huge statistical error of the vacuum-subtracted disconnected
correlator.
Using a noise reduction method and a large number of configurations
for the disconnected correlator,
we obtain the mass of $\sigma$ ($m_\sigma$) with a reasonable error.
From the results, we find that $\sigma$ is lighter than pion $\pi$, 
corresponding to the NG boson in the chiral broken phase, 
in all the fermion masses.
This peculiar feature is considered as
a reflection of the conformal dynamics of this theory.
While the light $\sigma$ in the $N_f = 12$ QCD
would not be regarded as a composite Higgs boson,
it is a promising signal for a walking technicolor
that has approximate conformal symmetry.
All the results in this report have already been
presented in Refs.~\cite{Aoki:2013pca,Aoki:2013zsa,Aoki:2013twa}.

\section{Simulation details}
We employ tree-level Symanzik gauge action
and HISQ (highly improved staggered quark)~\cite{Follana:2006rc} action without 
tadpole improvement and mass correction in the Naik term.
The flavor symmetry breaking of this action is highly suppressed 
in QCD ~\cite{Bazavov:2011nk} 
and we observed that it is almost negligible in our $N_f = 12$ 
QCD simulations~\cite{Aoki:2012eq}.
Using three degenerate staggered fermion species, 
we generate configurations by the standard HMC algorithm
on three different lattice volumes $(V=L^3)$,
$L=24, 30$ and $36$, with fixed aspect ratio $T/L=4/3$ at a single lattice 
spacing corresponding to $\beta \equiv 6/g^2=4.0$ for
the four fermion masses, $m_f=0.05, 0.06, 0.08$, and 0.10.
All the simulation parameters and also the corresponding $\pi$ mass ($m_\pi$)
are tabulated in Table~\ref{tab:table}.
We accumulate 8000--30000 trajectories depending on the parameters, and 
perform measurements every 2 trajectories.
Such a large number of configurations allows us 
to obtain a reasonable signal of $m_\sigma$. 
The statistical error is estimated by 
the standard jackknife method with bin size
of 200 trajectories.

We use the flavor-singlet scalar operator of
the local staggered fermion bilinear 
\begin{equation}
O_S(t) = \sum_{i=1}^3 \sum_x \bar{\chi}_i(x,t) \chi_i(x,t), 
  \label{eq:op}
\end{equation}
where $i$ denotes the index of the staggered fermion species.
Using $O_s(t)$ we calculate the flavor-singlet scalar correlator, 
which is constructed 
by both the connected $C(t)$ and vacuum-subtracted disconnected 
$D(t)$ correlators as
$\langle O_s(t) O_s^\dag(0) \rangle \propto 3 D(t) - C(t)$, 
where the factor in front of $D(t)$ comes from the number of species. 

In the staggered fermion formulation, the operator $O_s$ overlaps 
with $\sigma$ state, but also 
with a flavor non-singlet pseudoscalar state (${\pi_{\rm \overline{SC}}}$)
that is the staggered parity partner of $\sigma$. 
Therefore, in the large $t$ region, the correlator behaves as
\begin{equation}
  \label{eq:fermionic-combined}
3 D(t) - C(t) = A_\sigma(t) + (-1)^t A_{\pi_{\rm \overline{SC}}}(t),
\end{equation}
where $A_H(t) = A_H (e^{-m_H t} + e^{-m_H (T-t)})$, and 
the pseudoscalar state has a $(\gamma_5\gamma_4 \otimes \xi_5\xi_4)$ 
spin-taste structure, but is species-singlet. 

Since $C(t)$ can be regarded as a flavor non-singlet scalar correlator, 
$C(t)$ has a contribution from the lightest non-singlet scalar state ($a_0$) 
and its staggered parity partner ($\pi_{\rm SC}$). 
When $t$ is large, we can therefore write
\begin{equation}
  \label{eq:fermionic-connected}
 -C(t)  =  A_{a_0}(t) + (-1)^t A_{\pi_{\rm SC}}(t),
\end{equation}
where both $a_0$ and $\pi_{\rm SC}$ are species non-singlet 
and have the same taste structure as $\sigma$ and $\pi_{\rm \overline{SC}}$, 
respectively. 
The $\pi_{\rm SC}$ state is degenerate with the NG $\pi$ 
and also with $\pi_{\rm \overline{SC}}$,
($m_{\pi_{\rm SC}} = m_{\pi} = m_{\pi_{\rm \overline{SC}}}$) 
when the taste symmetry, thus the full flavor symmetry, is recovered.
From Eqs.~(\ref{eq:fermionic-combined}) and (\ref{eq:fermionic-connected}), 
the large $t$ asymptotic form of $3D(t)$ is written as 
\begin{equation}
  \label{eq:fermionic-disconnected-only}
  3 D(t)  =  A_{\sigma}(t) - A_{a_0}(t) + (-1)^t
(A_{\pi_{\rm SC}}(t) - A_{\pi_{\rm \overline{SC}}}(t)).
\end{equation}

The disconnected correlator $D(t)$, which is essential to 
obtain $m_\sigma$, is calculated by inverting the 
staggered Dirac operator at each spacetime point $(\vec{x},t)$. 
The computational cost of this inversion is mitigated by using 
a stochastic noise method. 
Moreover, its large fluctuations from the random noise in 
the method is dealt with by using a variance reduction 
method~\cite{Venkataraman:1997xi,Kilcup:1986dg,McNeile:2012xh} 
already employed for the flavor-singlet 
pseudoscalar~\cite{Venkataraman:1997xi,Gregory:2007ev} and 
chiral condensate~\cite{McNeile:2012xh} in usual QCD, 
and for the flavor-singlet scalar meson in $N_f = 12$ QCD~\cite{Jin:2012dw}
and in $N_f = 8$ QCD~\cite{Aoki:2013qxa}.
We employ $64$ random sources spread in the spacetime and the color space
for this reduction method.

 \begin{table}
\hfil
  \begin{tabular}{c|c|r|l|l|l}\hline
     $L^3 \times T$ & $m_f$ & \multicolumn{1}{|c}{$N_{\rm cfgs}$} & \multicolumn{1}{|c|}{$m_\sigma$} & \multicolumn{1}{c}{$m_\pi$} & \multicolumn{1}{|c}{$m_\sigma/m_\pi$}\\
     \hline
     $24^3 \times 32$ & 0.05 &11000 & 0.237(13)($^{02}_{01}$) & 0.3273(19)$^*$ & 0.73(4)($^{1}_{0}$) \\
     $24^3 \times 32$ & 0.06 & 14000 & 0.279(17)($^{07}_{01}$) & 0.3646(16)$^*$ & 0.77(5)($^{2}_{0}$) \\
     $24^3 \times 32$ & 0.08 & 15000 & 0.359(21)($^{01}_{18}$) & 0.4459(11)& 0.81(5)($^{0}_{4}$) \\
     $24^3 \times 32$ & 0.10 &  9000 & 0.453(42)($^{37}_{08}$) & 0.5210(7) & 0.87(8)($^{7}_{2}$) \\
     $30^3 \times 40$ & 0.05 & 10000 & 0.275(13)($^{21}_{08}$) &  0.3192(14)$^*$ & 0.86(4)($^{7}_{3}$)\\
     $30^3 \times 40$ & 0.06 & 15000 & 0.329(15)($^{47}_{12}$) & 0.3648(9)$^*$ & 0.90(4)($^{13}_{3}$)\\
     $30^3 \times 40$ & 0.08 & 15000 & 0.382(21)($^{03}_{16}$) & 0.4499(8) & 0.85(5)($^{1}_{4}$)\\
     $30^3 \times 40$ & 0.10 &  4000 & 0.431(51)($^{06}_{04}$) & 0.5243(7) & 0.82(10)($^{1}_{1}$)\\
     $36^3 \times 48$ & 0.05 &  5000 & 0.283(23)($^{01}_{02}$) & 0.3204(7)$^*$ & 0.88(7)($^{0}_{1}$)\\
     $36^3 \times 48$ & 0.06 &  6000 & 0.305(22)($^{25}_{06}$) & 0.3636(9)$^*$ & 0.84(6)($^{7}_{2}$)\\\hline
   \end{tabular}
   \caption{\label{tab:simulations} 
Parameters of lattice simulations, $m_\sigma$, and $m_\pi$.
$N_{\rm cfgs}$ is the number of saved gauge configurations. 
The second error of $m_\sigma$ is a systematic error coming from 
the fit range. 
The values of $m_\pi$ are from Ref.~\cite{Aoki:2012eq}, 
but the ones with ($^*$) have been updated. 
The error on $m_\sigma/m_\pi$ comes only from $m_\sigma$.}
\label{tab:table}
 \end{table}

\section{Result}

\begin{figure}[htbp]
\centering
   \includegraphics[width=6.5cm,clip]{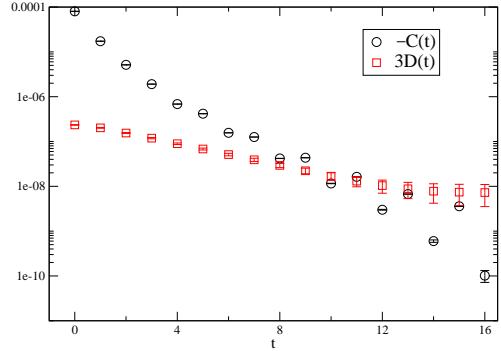}
 \caption{
Connected $-C(t)$ and vacuum-subtracted disconnected $3D(t)$ correlators 
on $L=24$ at $m_f=0.06$.
}
 \label{fig:corr}
\end{figure}

A typical result of $-C(t)$ and $3D(t)$ is shown in Fig.~\ref{fig:corr}.
In the large $t$ region, $3D(t)$ behaves as a smooth function of $t$ 
in contrast to $-C(t)$, which has a clear oscillating behavior. 
This means that the taste-symmetry breaking between 
$A_{\pi_{\rm SC}}(t)$ and $A_{\pi_{\rm \overline{SC}}}(t)$ 
in Eq.~(\ref{eq:fermionic-disconnected-only}) is small,
which can be expected from our previous work~\cite{Aoki:2012eq},
where the taste-symmetry breaking in the meson masses were negligible.

\begin{figure}[htbp]
\centering
   \includegraphics[width=6.5cm,clip]{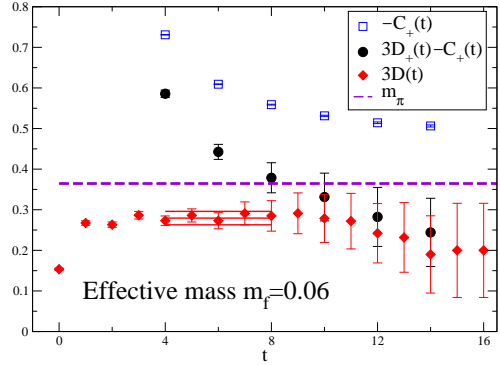}
 \caption{
Effective masses for parity-projected connected correlator 
$-C_+(t)$,
parity-projected full correlator $3D_+(t) - C_+(t)$,
and vacuum-subtracted disconnected correlator $3D(t)$
on $L=24$ at $m_f = 0.06$.
Dashed and solid lines express $m_\pi$ and fit result of $m_\sigma$
from $3D(t)$ in $t=4$--8 with the error band of one standard deviation,
respectively.
}
 \label{fig:meff}
\end{figure}

The effective mass of each correlator is presented in Fig.~\ref{fig:meff}.
In order to minimize the parity partner contributions, such as
$A_{\pi_{\rm \overline{SC}}}(t)$ in Eq.~(\ref{eq:fermionic-combined}),
we adopt a projection, $C_+(t) = 2C(t) + C(t+1) + C(t-1)$, at even $t$. 
The effective mass of the full correlator $3D_+(t)-C_+(t)$ at large $t$ 
is smaller than $m_\pi$, while the error is large. 
Since $m_\sigma < m_{a_0}$ as shown in the figure,
we also employ $D(t)$ to extract $m_{\sigma}$. 
The effective mass plateau of $D(t)$ is consistent with the one of 
$3D_+(t)-C_+(t)$ in the large $t$ region. 
Furthermore the plot clarifies the importance of using $D(t)$ to extract 
$m_{\sigma}$, because it performs better in identifying the lightest scalar 
state, even at small $t$ region. 
This might be caused by a reasonable cancellation among 
contributions from excited scalar states and the $a_0$ state in $D(t)$. 

\begin{figure}[tbp]
\centering
   \includegraphics[width=6.5cm,clip]{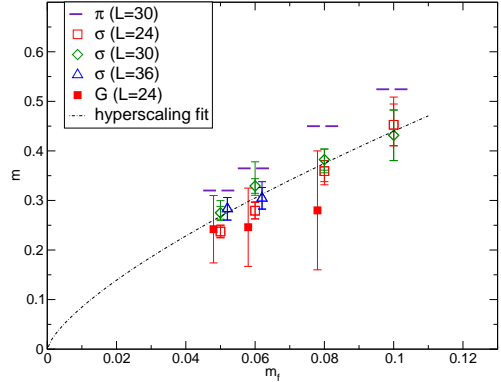}
\caption{
Mass of flavor-singlet scalar $m_\sigma$ in all the parameters
and mass obtained from gluonic calculation at $L=24$ are presented.
Inner error bar denotes statistical error and outer denotes combined
error of statistical and systematic added in quadrature.
Short dashed and dash-dot lines represent $m_\pi$ at each $m_f$
and hyperscaling fit result explained in text, respectively.
}
 \label{fig:mass}
\end{figure}

We fit $3D(t)$ between $t=4$ and $t=8$, assuming a single light state 
propagating in this region, to obtain $m_\sigma$ for all the parameters. 
A systematic error coming from the choice of the fitting range is 
estimated by the difference of central values obtained with 
several fit ranges. 
Figure~\ref{fig:mass} shows $m_\sigma$ of all the parameters
as a function of $m_f$, whose values are tabulated in 
Table.~\ref{tab:table}.
Finite size effects are negligible on the largest two volumes at each $m_f$
in our statistics.
The difference of $m_\sigma$ and $m_\pi$ is more than 
one standard deviation when the statistic and systematic errors 
are combined in quadrature, except for $m_f = 0.06$ on $L=30$, 
where there is a sizable systematic error.
Recently, the light $\sigma$ in the $N_f = 12$ QCD is also confirmed by
the other group with the different lattice action~\cite{Kuti:2013xyz}.
We also obtain the mass from 
the $0^{++}$ gluball correlators~\cite{Aoki:2013pca,Aoki:2013twa} with 
the variational analysis
at $L=24$ as plotted in the figure.
The result is smaller than $m_\pi$ by more than one standard deviation, 
and is consistent with $m_\sigma$ obtained 
from the fermionic correlator at each parameter,
while the statistical error is large.

\begin{figure}[tbp]
\centering
   \includegraphics[width=6.8cm,clip]{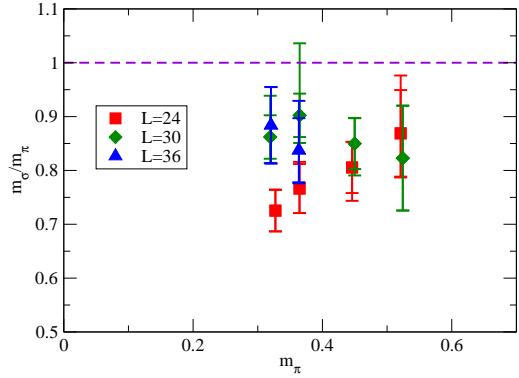}
\caption{
Ratio of $m_\sigma/m_\pi$ in all the parameters.
Inner error bar denotes statistical error and outer denotes combined
error of statistical and systematic added in quadrature.
}
 \label{fig:ratio}
\end{figure}

For a check of consistency with the hyperscaling of $m_\pi$
observed in the previous work~\cite{Aoki:2012eq}, 
we fit $m_\sigma$ on the largest volume data at each $m_f$ 
using the hyperscaling form 
\begin{equation}
m_{\sigma} = C m_f^{1/(1+\gamma)},
\end{equation}
with the fixed $\gamma = 0.414$ estimated from 
the finite-size hyperscaling analysis of $m_\pi$~\cite{Aoki:2012eq}.
This fit gives a reasonable value of $\chi^2/{\rm dof}= 0.12$,
and the fit result is shown in Fig.~\ref{fig:mass}. 
We also estimate the ratio $m_{\sigma}/m_{\pi}$ at each parameter 
and present it in Fig.~\ref{fig:ratio} and Table~\ref{tab:table}.
All the ratios are smaller than unity by more than one standard deviation 
including the systematic error, except the one at $m_f = 0.06$ on $L=30$, 
as previously explained. 
Since the ratio behaves as a constant for $m_f$,
this result is also shows that 
$m_f$ dependence of $m_\sigma$ is consistent with the one of $m_\pi$.
A constant fit with the largest volume data at each $m_f$ gives 0.86(3). 
These results are consistent with the theory being in the 
infrared conformal phase.

\section{Summary}

We have calculated the mass of the flavor-singlet scalar state in
$N_f = 12$ QCD with a HISQ-type action at $\beta = 4.0$.
We have obtained reasonable signals of $m_\sigma$ from 
the vacuum-subtracted disconnected correlator, which is
essential to calculate $\sigma$ state.
Clear signals in our simulations are possible thanks to the following 
salient features: 
1. Small taste-symmetry breaking, 
2. Efficient noise reduction methods, 
3. Large configuration ensembles, and 
4. Slow damping of $D(t)$ because of small $m_\sigma$.

From the results, we have confirmed that 
the obtained $m_\sigma$ is smaller than $m_\pi$ at all $m_f$
we simulated.
This feature is also observed from the gluonic measurements.
The hyperscaling behavior of $m_\sigma$ is confirmed from 
the analysis with a hyperscaling fit using the fixed value of $\gamma$,
which was estimated in our previous study~\cite{Aoki:2012eq}, 
and also from the ratio of $m_\sigma / m_\pi$.
These results are consistent with that the $N_f = 12$ QCD has
the infrared conformality.
Calculations with lighter fermion and at different lattice spacing
are important future works to understand more 
detailed property of $\sigma$ in this theory.

We regard the light $\sigma$ observed in the $N_f = 12$ QCD
as a reflection of the dilatonic nature of the conformal dynamics.
Thus, it is a promising signal for a walking theory~\cite{Yamawaki:1985zg}, 
which has an approximate infrared conformality.
It is the most pressing future direction to look at
more viable candidates for walking technicolor models.
For example, it will be interesting to investigate the scalar
spectrum of the $N_f=8$ QCD theory, which was shown to
be a good candidate for the walking technicolor model~\cite{Aoki:2013xza}.
We have already started this calculation and presented preliminary results
in this conference~\cite{Aoki:2013qxa}.

\section*{Acknowledgments}
Numerical calculations have been carried out
on the high-performance computing system $\varphi$ at KMI, Nagoya University, 
and the computer facilities of the Research Institute 
for Information Technology in Kyushu University.	
This work is supported by the JSPS Grant-in-Aid for Scientific Research 
(S) No.22224003, (C) No.23540300 (K.Y.), for Young Scientists (B) No.25800139 (H.O.) and No.25800138 (T.Y.), 
and also by Grants-in-Aid of the Japanese Ministry for Scientific Research on Innovative Areas No.23105708 (T.Y.). 
E.R. was supported by a SUPA Prize Studentship 
and a FY2012 JSPS Postdoctoral Fellowship for Foreign Researchers (short-term).

\bibliography{nf12_scalar}

\end{document}